\documentclass[twocolumn,showpacs,preprintnumbers,amsmath,amssymb,superscriptaddress,prl]{revtex4}
\usepackage{graphicx}
\usepackage{dcolumn}
\usepackage{bm}

\begin{document}

\title{Zero-bias anomaly in cotunneling transport through quantum-dot
spin valves}

\author{Ireneusz Weymann}
\affiliation{Department of Physics, Adam Mickiewicz University,
61-614 Pozna\'n, Poland}

\author{J\'ozef Barna\'s} \affiliation{Department of Physics, Adam Mickiewicz
University, 61-614 Pozna\'n, Poland} \affiliation{Institute of
Molecular Physics, Polish Academy of Sciences, 60-179 Pozna\'n,
Poland}

\author{J\"urgen K\"onig}
\affiliation{Institut f\"ur Theoretische Physik III,
Ruhr-Universit\"at Bochum, 44780 Bochum, Germany}

\author{Jan Martinek}
\affiliation{Institut f\"ur Theoretische Festk\"orperphysik,
Universit\"at Karlsruhe, 76128 Karlsruhe, Germany}
\affiliation{Institute for Materials Research, Tohoku University,
Sendai 980-8577, Japan} \affiliation{Institute of Molecular
Physics, Polish Academy of Sciences, 60-179 Pozna\'n, Poland}

\author{Gerd Sch\"on} \affiliation{Institut f\"ur Theoretische
Festk\"orperphysik, Universit\"at Karlsruhe, 76128 Karlsruhe,
Germany}

\date{\today}

\begin{abstract}

We predict a new zero-bias anomaly in the differential conductance
through a quantum dot coupled to two ferromagnetic leads with
antiparallel magnetization. The anomaly differs in origin and
properties from other anomalies in transport through quantum dots,
such as the Kondo effect. It occurs in Coulomb-blockade valleys
with an unpaired dot electron. It is a consequence of the
interplay of single- and double-barrier cotunneling processes and
their effect on the spin accumulation in the dot. The anomaly
becomes significantly modified when a magnetic field is applied.
\end{abstract}


\maketitle

{\it Introduction.} -- The combination of Coulomb interaction
effects, which frequently are strong in nanostructures, and
spin-dependent transport in systems coupled to ferromagnetic leads
opens a new field of research with qualitatively new transport
properties \cite{ono97,barnas98}. Spin-dependent transport through
nonmagnetic grains may be influenced by the presence of spin
accumulation \cite{chye02} leading to a different transmission for
parallel and antiparallel orientation of the leads' magnetization,
which results in a finite tunnel magnetoresistance \cite{bulka00}.
In the limit of weak dot-lead coupling and when a dot level is in
resonance with the Fermi level of the leads (linear response
regime), transport is dominated by sequential tunneling. Away from
resonance, sequential tunneling is exponentially suppressed, and
transport is due to higher-order tunneling
\cite{averin90,cot_exp}. In the Coulomb-blockade valley with an
unpaired electron occupying the dot, Kondo-assisted tunneling
\cite{kondo} gives rise to a pronounced zero-bias anomaly in the
differential conductance at temperatures below the Kondo
temperature $T_{\rm K}$. Above $T_{\rm K}$, transport is dominated
by (second-order) cotunneling, with regular zero-bias behavior for
nonmagnetic leads.

In this paper, we study cotunneling transport through a
single-level and singly-occupied quantum dot attached to
ferromagnetic leads. When source and drain electrodes are
magnetized antiparallel to each other, we find a pronounced
zero-bias anomaly that is completely unrelated to Kondo
correlations. It is rather a consequence of the interplay of spin
accumulation and spin relaxation due to spin-flip cotunneling. A
finite spin accumulation on the quantum dot partially suppresses
transport. Spin-flip cotunneling provides a channel of spin
relaxation and, hence, reduces the spin accumulation. As we show
below, single-barrier spin-flip cotunneling (in the absence of a
magnetic field) plays a role in linear response, $|eV| \ll k_{\rm
B}T$, but is negligible in the opposite limit. This gives rise to
a zero-bias anomaly in the differential conductance. The
prediction of this zero-bias anomaly as well as the study of its
properties is the central issue of this article.

{\it Model and Method.} -- We consider a quantum dot with a single
level at energy $\varepsilon$ coupled to ferromagnetic leads with
either parallel or antiparallel magnetization directions. The
model Hamiltonian is $H=H_{\rm L}+H_{\rm R}+H_{\rm D}+H_{\rm T}$.
The terms $H_{r}=\sum_{q\sigma} \varepsilon_{r
q\sigma}c^{\dagger}_{r q\sigma} c_{r q\sigma}$ for $r = {\rm
L},{\rm R}$ represent noninteracting electrons in the left and
right lead, where $\varepsilon_{r q\sigma}$ denotes the energy of
an electron with wave number $q$ and spin $\sigma$ in lead $r$.
The dot is modelled by $H_{\rm D}=\sum_{\sigma} (\varepsilon
\pm\Delta/2) d^{\dagger}_{\sigma} d_{\sigma}+U
d^{\dagger}_{\uparrow} d_{\uparrow}
d^{\dagger}_{\downarrow}d_{\downarrow}$, where $\Delta$ is the
Zeeman energy due to an external magnetic field, $U$ is the
charging energy, and the $+$ $(-)$ sign corresponds to $\sigma
=\uparrow (\downarrow )$. Tunneling between the dot and leads is
described by $H_{\rm T}=\sum_{r q \sigma} ( t_r c^{\dagger}_{r q
\sigma} d_{\sigma} +t^{*}_r d_{\sigma}^{\dagger} c_{r q \sigma}
)$. Ferromagnetism of the leads is included {\it via}
spin-dependent densities of states, $\rho_{r}^{\uparrow} \neq
\rho_{r}^{\downarrow}$. The degree of spin polarization in the
leads is characterized by the factor $p_{r}=(\rho_{r}^{+} -
\rho_{r}^{-})/ (\rho_{r}^{+}+\rho_{r}^{-})$, where $\rho_{r}^{+}$
$(\rho_{r}^{-})$ is the density of states for spin-majority
(spin-minority) electrons. The tunnel-coupling strength is
characterized by $\Gamma_{r}^{\sigma}= 2\pi |t_{r}|^2
\rho_{r}^{\sigma}$. Finally, we define $\Gamma_{r} \equiv
(\Gamma_{r}^{\uparrow} + \Gamma_{r}^{\downarrow})/2$,
$\Gamma^\sigma \equiv \Gamma_{\rm L}^\sigma + \Gamma_{\rm
R}^\sigma$, and $\Gamma \equiv \Gamma_{\rm L} + \Gamma_{\rm R}$,
and assume $\Gamma_{\rm L}=\Gamma_{\rm R}\equiv \Gamma/2$.

We consider the Coulomb-blockade valley in which the dot is singly
occupied with either spin. The probabilities for occupation with
spin $\sigma$ are $P_\sigma$. We determine the rate
$\gamma^{\sigma \Rightarrow \sigma'}_{rr'}$ for a cotunneling
process, in which one electron leaves the dot to reservoir $r'$
and one electron enters from $r$ with the initial and final dot
state being $\sigma$ and $\sigma'$, respectively, in second-order
perturbation theory. For $\sigma=\sigma'$, i.e., when the dot spin
is not changed, and $\Delta=0$, the corresponding rate is
\cite{remark}
\begin{eqnarray}
  \gamma^{\sigma \Rightarrow \sigma}_{rr'} &=&
  \frac{1}{2\pi\hbar}
  {\rm Re} \int d\omega [1-f(\omega-\mu_r)]f(\omega-\mu_{r'}) \times
  \nonumber \\
  &&
  \left[ \frac{\Gamma^\sigma_r \Gamma^{\sigma}_{r'}}
       {(\omega-\varepsilon+i0^+)^2} +
  \frac{\Gamma^{\bar \sigma}_r \Gamma^{\bar \sigma}_{r'}}
       {(\omega-\varepsilon-U+i0^+)^2}
    \right] \;,
\end{eqnarray}
while we get
\begin{eqnarray}
  \gamma^{\sigma \Rightarrow \bar \sigma}_{rr'} &=&
  \frac{\Gamma^\sigma_r \Gamma^{\bar \sigma}_{r'}}{2\pi\hbar}
  {\rm Re} \int d\omega [1-f(\omega-\mu_r)]f(\omega-\mu_{r'}) \times
\nonumber \\
  &&
  \left[ \frac{1}{\omega-\varepsilon+i0^+} + \frac{1}{\varepsilon+U-\omega+i0^+}
    \right]^2 \;,
\end{eqnarray}
for cotunneling process in which the dot spin is flipped
($\bar\sigma$ is the opposite spin of $\sigma$). Here,
$f(\omega-\mu_r)$ is the Fermi function of reservoir $r$ with
electro-chemical potential $\mu_r$.

The probabilities $P_\sigma$ are obtained from the stationary rate
equation $0=\sum_{rr'} \left[ P_\uparrow \gamma^{\uparrow
\Rightarrow \downarrow}_{rr'} - P_\downarrow \gamma^{\downarrow
\Rightarrow \uparrow}_{rr'}\right] $ together with the
normalization condition $P_\uparrow + P_\downarrow =1$. The
current $I$ is, then, given by
\begin{equation}
  I = e \sum_{\sigma\sigma'} P_\sigma \left[
  \gamma^{\sigma \Rightarrow \sigma'}_{\rm LR} -
  \gamma^{\sigma \Rightarrow \sigma'}_{\rm RL} \right]
  \;.
\end{equation}

\begin{figure}[tb]
\includegraphics[width=0.95\columnwidth]{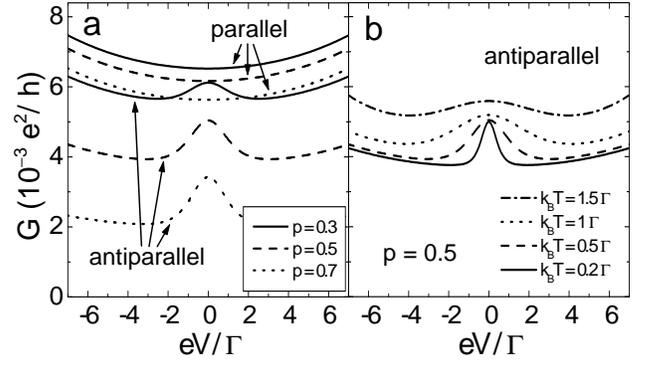}
\caption{\label{Fig1}
  (a) Differential conductance in the parallel and antiparallel configurations
  as a function of bias voltage for different values of spin polarization $p$
  at $\varepsilon =-U/2$, $U=30 \Gamma$, $k_{\rm B}T=0.5 \Gamma$, and (b) for
  different temperatures and $p=0.5$.}
\end{figure}

{\it Results in the absence of magnetic field.} -- We consider
symmetrically polarized leads, $p_{\rm L}=p_{\rm R}\equiv p$ and,
first, ignore a Zeeman splitting, $\Delta = 0$. The differential
conductance $G = \partial I/\partial V \equiv (e^2/h) g$ as a
function of the bias voltage is shown in Fig.~\ref{Fig1}(a). For
leads magnetized in parallel, we find the typical parabolic
behavior of the cotunneling conductance with increasing bias
voltage. This is distinctively different for the antiparallel
configuration, for which a zero-bias anomaly appears. The
bias-voltage dependence of the conductance for different
temperatures is shown in Fig.~\ref{Fig1}(b), where the width of
the peak grows with $T$. It is possible to get some insight into
the basic properties of this behavior using analytical results
obtained in a limit in which the formulas simplify considerably
while all the main physics of the zero-bias anomaly remains
included. Deep inside the Coulomb-blockade regime, we can neglect
corrections in the ratios $A/B$ with $A= |eV|, k_{\rm B}T$ and
$B=|\varepsilon|,\varepsilon+U$. We find
\begin{equation}
  g^{\rm P} = \frac{\Gamma^2}{2} \left[
    \frac{1}{\varepsilon^2} + \frac{1}{(\varepsilon +U)^2}
    + \frac{1-p^2}{|\varepsilon| (\varepsilon +U)} \right]
\end{equation}
for the parallel configuration, independent of $|eV|/k_{\rm B}T$.
For the antiparallel configuration, we get
\begin{equation}
  g^{\rm AP}_{\rm max} =   \frac{\Gamma^2}{2} (1-p^2) \left[
    \frac{1}{\varepsilon^2} + \frac{1}{(\varepsilon +U)^2}
    + \frac{1}{|\varepsilon| (\varepsilon +U)} \right]
\end{equation}
in linear response, $|eV| \ll k_{\rm B}T$, and
\begin{equation}
  g^{\rm AP}_{\rm min} =   \frac{\Gamma^2}{2} \frac{1-p^2}{1+p^2} \left[
    \frac{1}{\varepsilon^2} + \frac{1}{(\varepsilon +U)^2}
    + \frac{1-p^2}{|\varepsilon| (\varepsilon +U)} \right]
\end{equation}
for $|eV| \gg k_{\rm B}T$. We see that $g^{\rm P} > g^{\rm AP}$,
i.e., the system displays a tunnel magnetoresistance. But,
furthermore, we find a zero-bias peak, since $g^{\rm AP}_{\rm max}
> g^{\rm AP}_{\rm min}$, whose relative strength, characterized by
$x \equiv (g_{\rm max}^{\rm AP} - g_{\rm min}^{\rm AP})/g_{\rm
min}^{\rm AP}$, increases from $x=p^2$ at the edges
($|\varepsilon| \ll \varepsilon + U$ or $|\varepsilon| \gg
\varepsilon + U$) to $x= 4p^2/(3-p^2)$ in the middle ($\varepsilon
= -U/2$) of the Coulomb-blockade valley.

The zero-bias anomaly, present (for $\Delta=0$) only in the
antiparallel configuration, has the following properties:\\
(i) The crossover from $g^{\rm AP}_{\rm max}$ to $g^{\rm AP}_{\rm
min}$ is around $|eV| \approx k_{\rm B}T \sqrt{8/(1+p^2)}$, i.e.,
the width of the zero-bias anomaly scales linearly with $k_{\rm
B}T$ and depends only weakly on $p$.\\
(ii) The {\em relative} peak height $x$ increases monotonically
with $p$ and when moving from the edges towards the middle of the
Coulomb-blockade valley.\\
(iii) The {\em absolute} peak height $g_{\rm max}^{\rm AP} -
g_{\rm min}^{\rm AP}$ depends nonmonotonically on $p$, since it
vanishes for $p=0$ and $p=1$.\\
(iv) At low temperature both $g_{\rm max}^{\rm AP}$ and $g_{\rm
min}^{\rm AP}$ increase with temperature, $g_{\rm max, min}^{\rm
AP}(T) / g_{\rm max, min}^{\rm AP}(0) = 1+(T/B)^2+{\cal O}(T^4)$
with the same constant $B$, such that $x$ is nearly independent of
temperature.

Processes responsible for the zero-bias anomaly in the cotunneling
regime are of second order, while these leading to the Kondo
effect are of higher than second order.  The zero-bias anomaly of
the cotunneling current is therefore distinctively different from
that associated with the Kondo effect. The latter occurs at low
temperature, $T \lesssim T_{\rm K}$, shows up in the parallel
configuration as well \cite{martinek03,pasupathy04}, grows
logarithmically with decreasing temperature, and reaches perfect
transmission, $g=1$. Its width at low temperature saturates at
$k_{\rm B} T_{\rm K}$, and it has a different magnetic-field
dependence.

We close with the remark that the exchange field due to the
presence of ferromagnetic leads discussed in
Ref.~\onlinecite{martinek03,koenig03} does not affect transport in
the case considered here. Spin precession does not appear since
the leads are magnetized collinearly.

\begin{figure}[t]
\includegraphics[width=0.87\columnwidth,clip]{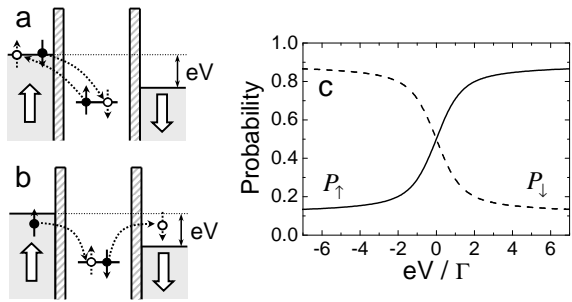}
\caption{\label{Fig2}
  Single-barrier (a) and double-barrier (b) cotunneling processes,
  and the occupation probabilities for spin-up and spin-down
  electrons in the antiparallel configuration (c).
  The parameters are $k_{\rm B}T= 0.5 \Gamma$, $U=30 \Gamma$,
  $\varepsilon =-U/2$, and $p=0.5$.
  }
\end{figure}

{\it Mechanism of the zero-bias anomaly.} -- To understand the
mechanism of the zero-bias anomaly we distinguish between four
different types of cotunneling processes. In each of them two
tunneling events are involved, either through the same or through
the two opposite tunnel barriers. Accordingly, we refer to them as
single-barrier [Fig.~\ref{Fig2}(a)] and double-barrier cotunneling
[Fig.~\ref{Fig2}(b)]. Furthermore, the two electrons involved may
carry the same or opposite spin, i.e., both single-barrier and
double-barrier events come as either spin-flip or non-spin-flip
cotunneling. In calculation we have taken into account all
possible cotunneling processes. Here, however, we discuss just the
ones responsible for the anomaly. Double-barrier cotunneling
contributes directly to the current, while single-barrier
cotunneling preserves the total charge. Nevertheless, spin-flip
single-barrier cotunneling can influence the total current
indirectly, by changing of the magnetic state of the dot.
In the antiparallel configuration, the dot hosts a nonequilibrium
spin accumulation $m=(P_\uparrow - P_\downarrow)/2$. A different
occupation of up- and down-spin levels in the dot, $P_\uparrow
\neq P_\downarrow$, appears (even for $\Delta=0$) when the
spin-flip cotunneling rates that change the dot from $\uparrow$ to
$\downarrow$ and $\downarrow$ to $\uparrow$ are different from
each other. In equilibrium, $V=0$, both rates are trivially the
same and, hence, $P_\uparrow =P_\downarrow$. The situation is
different at finite bias voltage and antiparallel magnetized
electrodes. Now, only the two spin-flip processes that transfer an
electron from the left to the right lead determine the magnetic
state of the dot. The one shown in Fig.~\ref{Fig2}(b) changes the
dot spin from $\downarrow$ to $\uparrow$. Since only majority
spins of the electrodes are involved, the corresponding rate is
larger than that of the other process that changes the dot spin
from $\uparrow$ to $\downarrow$ by using minority spins only. This
results in a nonequilibrium spin accumulation $m > 0 $
($P_\uparrow > P_\downarrow$) that increases with $V$
[Fig.~\ref{Fig2}(c)]. The initial state for the dominant spin-flip
cotunneling process that contributes to the current,
Fig.~\ref{Fig2}(b), is $\downarrow$. Thus, the reduced probability
$P_\downarrow$ decreases transport. This is the mechanism by which
spin accumulation gives rise to the tunnel magnetoresistance
effect, $g^{\rm P} > g^{\rm AP}$.

Any spin-flip process, that reduces the spin accumulation will
enhance the conductance. Such a process is provided by
single-barrier spin-flip cotunneling. The corresponding rate
scales with $k_{\rm B} T$ while that of double-barrier cotunneling
is proportional to $\max\{|eV|, k_{\rm B}T \}$. This explains the
zero-bias anomaly: For $|eV| \lesssim k_{\rm B}T$, single-barrier
processes play a significant role, and, therefore, the current
increases relatively fast with applied bias, which yields $g^{\rm
AP}_{\rm max}$. For $|eV| \gg k_{\rm B}T$, on the other hand, the
relative role of single-barrier spin-flip processes is negligible
as compared to double-barrier cotunneling, and the conductance is
reduced to $g^{\rm AP}_{\rm min}$. It is, thus, the interplay of
spin-dependent single- and double barrier cotunneling processes
that gives rise to the zero-bias anomaly in the differential
conductance. This zero-bias anomaly is distinctively different
from experimentally-observed peaks in the differential conductance
for nonmagnetic systems \cite{cot_exp} which occur at the onset of
sequential tunneling.

Finally, we remark that no zero-bias anomaly occurs in the
Coulomb-blockade valleys with an even number ($0$ for $\varepsilon
> 0$ and $2$ for $\varepsilon+U < 0$), as in this case the total
dot spin is zero, and spin accumulation is absent.

\begin{figure}[tb]
  \includegraphics[width=0.95\columnwidth]{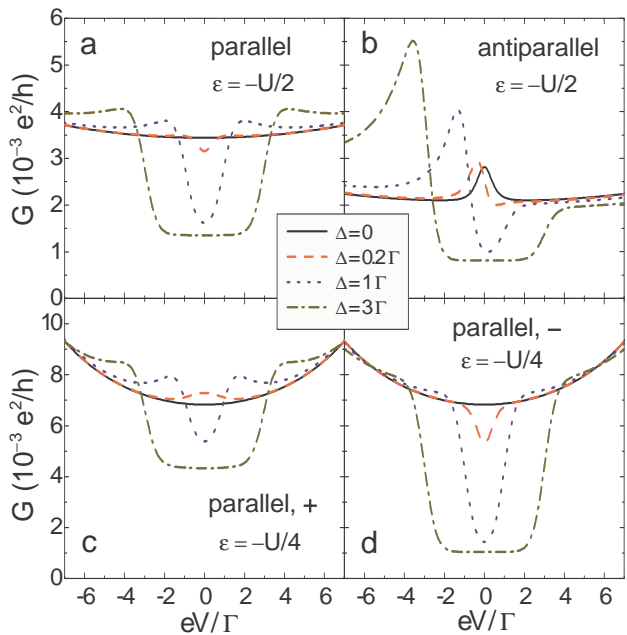}
  \caption{\label{Fig3}
    (color online) Differential conductance in presence of a Zeeman splitting
    with $p=0.5$, $k_{\rm B}T = 0.2 \Gamma$, and $U=40 \Gamma$ for the
    symmetric (a,b) and asymmetric (c,d) Anderson model with parallel (a,c,d)
    and antiparallel (b) relative magnetization of the leads.
    In (c), the Zeeman splitting favors a dot polarization parallel,
    in (d) antiparallel to the leads.}
\end{figure}

{\it Results in the presence of magnetic field.} -- In the
presence of an external magnetic field, the dot levels are split
by a Zeeman energy $\Delta$. We restrict ourselves to the case of
an external field that is collinear with the leads' magnetization
directions. When the Zeeman splitting $\Delta$ is larger than both
temperature and bias voltage, $|\Delta| \gg \max \{ k_{\rm B}T,
|eV| \}$, only the lower spin level is occupied, i.e., the dot is
fully polarized. In this case, spin-flip cotunneling is completely
suppressed, which leads to a reduction of the conductance. With
the same approximations as we used for $\Delta =0$, we obtain
\begin{equation}
  g^{{\rm P},\pm}_{\rm field} = \frac{\Gamma^2}{4} \left[
    \frac{(1\pm p)^2}{(\varepsilon-|\Delta|/2)^2} +
    \frac{(1\mp p)^2}{(\varepsilon+U+|\Delta|/2)^2} \right]
\end{equation}
for the parallel configuration. Here, $+/-$ correspond to the
cases when the Zeeman splitting favors a dot polarization that is
parallel or antiparallel to the leads, respectively. We remark
that for $\varepsilon = -U/2$ the conductance is symmetric under
reversal of the magnetic field, in contrast  to $\varepsilon \neq
-U/2$, were the field reversal results in a different conductance
[Fig.~\ref{Fig3}(c,d)],  similar to the spin-readout scheme
proposed in Ref.~\onlinecite{recher00}. For antiparallel
magnetized leads, we find
\begin{equation}
  g^{\rm AP}_{\rm field} = \frac{\Gamma^2}{4} (1-p^2) \! \left[
    \frac{1}{(\varepsilon \! - \! |\Delta|/2)^2} \! + \!
    \frac{1}{(\varepsilon \! + \! U \! + \! |\Delta|/2)^2} \right] .
\end{equation}
The above expressions approximate the plateaus shown in
Fig.~\ref{Fig3}. When the bias voltage is increased such that
$|eV| \sim |\Delta| \gg k_{\rm B}T$, spin-flip processes are again
possible, and the dot is no longer fully polarized. For
antiparallel magnetization of the leads and fixed orientation of
the external field, the conductance is asymmetric under reversal
of bias voltage, see Fig.~\ref{Fig3}(b). The reason is that
single-barrier spin-flip cotunneling favors the dot state with the
lower energy, which is, depending on the direction of the voltage
drop, either the right or the wrong initial state for the dominant
double-barrier cotunneling contribution to the current. The
magnetic-field dependence of the peak height reflects the fact
that single-barrier spin-flip cotunneling increases linearly with
$\Delta$.

In the parallel configuration, a plateau evolves for $\max \{
k_{\rm B}T, |eV| \} \ll |\Delta|$, again due to suppression of the
spin-flip contributions to the current. In addition, we find that
a zero-bias anomaly evolves even for parallel leads, when the
Zeeman splitting favors a dot polarization in the same direction,
Fig.~\ref{Fig3}(c). As the dominant transport channel is
non-spin-flip cotunneling with majority spins, and for $|eV| \ll
|\Delta| \lesssim k_{\rm B}T$, thermal occupation of the dot's
magnetic states favors occupation with a majority spin, transport
is enhanced as compared to $\Delta=0$.

{\it Summary.} -- We predict a zero-bias anomaly in cotunneling
transport through quantum dots attached to ferromagnetic leads
that are magnetized antiparallel to each other. This zero-bias
anomaly originates from the interplay of single-barrier and
double-barrier spin-flip cotunneling processes. From an
experimental point of view, the anomaly may be observed in quantum
dots and/or molecules attached to ferromagnetic leads, which
include an odd number of electrons. Such structures have been
already realized experimentally \cite{chye02,cot_exp,pasupathy04}.

We thank M. Braun, S. Maekawa, Yu.V. Nazarov, D. Ralph, and Y.
Utsumi for helpful discussions. The work was supported by the
projects PBZ/KBN/044/P03/2001 and 2 P03B 116 25, "Spintronics" RT
Network of the EC RTN2-2001-00440,  EC G5MA-CT-2002-04049, the DFG
through SFB 491, GRK 726, and under CFN.


\end{document}